# A LOW-ENERGY FAST CYBER FORAGING MECHANISM FOR MOBILE DEVICES

Somayeh Kafaie[1], Omid Kashefi[1] and Mohsen Sharifi[1]

[1]School of Computer Engineering,
Iran University of Science and Technology, Tehran, Iran
so_kafaie@comp.iust.ac.ir, kashefi@{ieee.org,iust.ac.ir},
msharifi@iust.ac.ir

## ABSTRACT

*The ever increasing demands for using resource-constrained mobile devices for running more resource intensive applications nowadays has initiated the development of cyber foraging solutions that offload parts or whole computational intensive tasks to more powerful surrogate stationary computers and run them on behalf of mobile devices as required. The choice of proper mix of mobile devices and surrogates has remained an unresolved challenge though. In this paper, we propose a new decision-making mechanism for cyber foraging systems to select the best locations to run an application, based on context metrics such as the specifications of surrogates, the specifications of mobile devices, application specification, and communication network specification. Experimental results show faster response time and lower energy consumption of benched applications compared to when applications run wholly on mobile devices and when applications are offloaded to surrogates blindly for execution.*

## KEYWORDS

*Pervasive Computing, Cyber Foraging, Offloading, Surrogate, Mobile Computing, Resource Constraint.*

## 1. INTRODUCTION

Nowadays mobile devices are very popular and people all over the world are increasingly using mobile devices such as cell phones and PDAs to run many applications from daily tasks to emergencies. Finally, using mobile devices and wireless networks, accessing information anywhere and anytime seems more achievable [1]. In recent years, users benefit from mobile devices to use more resource intensive applications. Some examples of such applications are natural language translator [2, 3], speech recognizer [2, 3], optical character recognizer [2], image processor [4], and games with high amount of processing [5].

However, there are often shortcomings in quality of mobile devices' tasks due to their resource poverty. The mentioned applications require higher computing power, memory, and battery lifetime than is available on resource constrained mobile devices. They also require faster responses than is currently supported on mobile devices. Unfortunately, at any level of cost and technology, considerations such as weight, size, battery life, ergonomics, and heat dissipation impose severe restrictions on computational resources such as processor speed, memory size and disk capacity of these devices [6]. Therefore, mobile devices always remain more resource constrained than traditional stationary computers [6, 7].

On the other hand, a pervasive computing environment is an environment that focuses on mobility and usage of mobile devices [8]. Pervasive computing was first introduced by Mark Weiser [9] in 1991; Satyanarayanan [10] has defined pervasive environments as "environments saturated with computing and communication capability, yet gracefully integrated with human users".

One of the most important and favourable solutions to cope with resource poverty of mobile devices, especially in pervasive computing, is *cyber foraging*. Generally, cyber foraging is task offloading in order to resource augmentation of a wireless mobile device by exploiting available static computers [10]. In cyber foraging approach, the mobile device sends the whole or a part of an application to nearby idle static computers, called *surrogate* and receives the results to improve the response time and/or accuracy, or confront with its resource constraint. In this paper, we study effectiveness of cyber foraging from mobile devices, surrogates, application, and network aspects.

The remainder of the paper is organized as follows. Related researches on task offloading are discussed in Section 2. In Section 3, the mobility constraints, cyber foraging idea and effectiveness of this idea to alleviate the constraints are explained. Section 4 presents our proposed cyber foraging approach. The results of experimental evaluations are depicted and discussed in Section 5, and Section 6 concludes the paper.

## 2. RELATED WORK

There are several approaches with different objectives that have used the offloading of applications, but the term "cyber foraging" was first introduced by Satyanarayanan [10]. Cyber foraging is the discovery of static idle computers called surrogates in the vicinity of a mobile device and entrusting some of the tasks of the mobile device to them [10]. As computers become cheaper and more plentiful, cyber foraging approaches become more reasonable to employ.

Spectra [3] is the first cyber foraging system that is focused on reducing the latency and energy consumption. Spectra adds a feature called *self-tuning* to monitor application behaviour and estimate the resource demand of an application. Spectra's approach to measure energy consumption of the tasks does not work well enough, in some cases. Furthermore developers must follow most of the cyber foraging steps in Spectra manually that it causes significant changes in the code.

Chroma [2, 11, 12] is an extension of Spectra which tries to improve it by reducing the burden on developers. To do so, Chroma uses a new concept called *tactics* that are meaningful ways of application partitioning, specified by the programmer. Chroma uses a fixed utility function to improve latency but ignores battery lifetime. Furthermore, Chroma presents three ways applicable in environments that are full of idle computing resources. First it sends a task execution request to several surrogates in parallel and chooses the fastest response; second it splits operation data and forwards each part to a different surrogate; third it sends the same task execution request with different quality to different surrogates and picks the result with the highest quality that satisfies the latency threshold.

On the other hand, Gu *et al.* [13] have used a graph model to select offloading parts of the program to improve memory constraint of mobile device. Ou *et al.* [14, 15] have expanded their approach and have used a similar method to address the CPU and bandwidth constraint, too. Song *et al.* [16, 17] has proposed a middleware architecture, called MobiGo, for seamless mobility to choose the best available service according to the bandwidth and latency, and Kristensen [18, 19] has introduced Scavenger as a cyber foraging framework whose focus is on CPU power of mobile devices.

However, none of the mentioned works, except Spectra, address directly energy constraint in mobile devices. Othrnan *et al.* [20] were one of the oldest researchers who employ the offloading to reduce the power consumption. Kemp *et al.* [21] also presented Ibis to compare offloading with local execution in terms of responsiveness, accuracy and energy consumption. Cuervo *et al.* [22] present an infrastructure, called MAUI to offload the applications and reduce the energy consumption. MAUI supports programs written in managed code environments such as Microsoft .Net CLR and Java. It provides a graph of program's methods and divides them into local and remote groups to execute. They have located the solver (decision-making unit) out of the mobile device to decrease the computation cost, while burden more communication cost.

In this paper, we propose a context-aware decision-making mechanism to make decisions about task offloading in terms of not only energy consumption, but also current processing power and available memory to improve response time and energy consumption in mobile devices.

## 3. CYBER FORAGING AND MOBILE COMPUTING

### 3.1. Augmented Mobile Devices

Mobile devices, due to their mobility nature, cannot be plugged in most of times. Therefore, energy consumption is one of the most important constraints of mobile devices [23]. On the other hand, portability requirements necessitate being as light and small as possible. The inherent constraints include low processor speed, memory, storage size, network bandwidth and limited battery lifetime.

Ubiquitous availability of advanced mobile technologies makes users to expect to run the same applications on mobile devices and static computers. However, regarding resource poverty of mobile devices, it is evident that static computers perform the tasks faster and more accurate. Besides, it is possible that the mobile device does not have sufficient memory, storage or battery to complete the task.

To run the task on a static computer (i.e. surrogate) on behalf of the mobile device, it is required to send the related code and data from the mobile device to the surrogate and receive back the results, which is a time and energy-consuming process. The time of sending/receiving data (application code, input parameters and results) to/from the surrogate depends on the size of data and results as well as on the network bandwidth.

Cyber foraging causes reduction of execution time and energy consumption due to the exploiting more powerful surrogates, but transmission of associated information increases response time and decreases battery lifetime. Since communication usually consumes more energy than computation [21], it raises an issue: "under which circumstances is it worth to use offloading?". Therefore, a decision system must imply that a task is worth to offload to a surrogate or not. In this paper, we present a mechanism to decide about task offloading according to the context information.

### 3.2. Cyber Foraging Steps

A cyber foraging approach includes some steps that every available cyber foraging systems have considered all or some of them. These steps can be summarized as follows.

- *Surrogate discovery.* First of all, available idle surrogates that are ready to share their resources with the mobile device must be found. Some researches [13, 18] have addressed surrogate discovery.

- *Context gathering*. To have a good decision about target execution location, there is a need to monitor available resources in surrogates and mobile devices and estimate application resource consumptions which is considered as context gathering in some cyber foraging systems [2, 3, 18].

- *Partitioning*. In this step, a task is divided into smaller size subtasks, and undividable i.e. unmovable parts are specified. Some researches [24] do the partitioning automatically.

- *Scheduling.* The most important step of cyber foraging is to place each task at the surrogate(s) or the mobile device most capable of performing it, based on the context information and the estimated cost of doing so. Many researches [3, 12, 13, 15, 19, 22] have considered this step. foraging is making

- *Remote execution control*. The final step involves the establishment of a reliable connection between the mobile device and the appropriate surrogate to pass required information, remote execution, and the receipt of returned results. Various researches [3, 4, 6, 12, 19] have considered remote execution control.

In this paper, we focus on *scheduling* step of cyber foraging and propose a decision-making mechanism to select the best location to run a mobile device's task according to the pre-gathered context information.

### 3.3. Cyber Foraging Goals

Cyber foraging is a solution to execute resource intensive applications on resource constrained mobile devices. In fact, available researches in cyber foraging have tried to augment some resources of mobile devices in terms of effective metrics to achieve more efficient application execution. The most important resources have been considered by offloading approaches are as follows:

- *Energy*. One of the most important constraints of mobile devices is energy consumption because mobile device's energy cannot be replenished by itself [23]. Many researches [3, 20-22] have considered energy consumption as a parameter for offloading

- *Memory and storage*. Memory capacity of mobile devices is less than stationary computers and memory intensive applications cannot usually run on mobile devices. Many researches [13-15] have considered the availability of memory and storage as another effective parameter for offloading decision.

- *Response time*. When the processing power of mobile devices is considerably lower than static computers, task offloading is beneficial to decrease execution time. There are many researches [3, 12, 14, 15, 19] that have considered the response time and latency as a major parameter affecting the offloading decision.

- *I/O*. Displaying a movie on a bigger screen, playing music on more powerful speakers, and printing are examples of task offloading to improve I/O quality or exploit more I/O devices. Some researches [16, 17] have focused on augmenting I/O as an effective parameter for offloading decision.

In this paper, we focus on energy, response time and memory. We offload the mobile device's tasks to decrease energy consumption and response time in mobile devices. Furthermore we consider memory demand of the task and available memory of every location (i.e. the mobile device and surrogates) to select appropriate location to execute the task.

## 4. PROPOSED DECISION MECHANISM

In this section we propose an approach to raise the participation rate of mobile devices in pervasive and mobile computing using context aware task offloading. The pervasive computing environment in our experiments includes a mobile device and some desktop computers as surrogates that are intra-connected through a wireless LAN.

When a task is requested to run on the mobile device, a solver program runs immediately to make a decision according to the context metrics either to offload whole the task or to execute it on mobile device itself.

### 4.1. Context Metrics

Due to the dynamic nature of resources involved in typical computational pervasive environments and portability of mobile devices, the ability of a device to perform the operations varies over time. Therefore, making decision to offload a task must be according to the current situation. We categorize the context metrics into four classes:

- *Mobile device metrics* include current processing power, available memory, and available energy.
- *Surrogate metrics* include current processing power, and available memory.
- *Network metrics* include network type and its current conditions that can change depending on the location such as data transmission rate and signal strength.
- *Application metrics* include application type, which is one of CPU intensive, memory intensive, and I/O intensive [25], and the size of application's code, input and output data. Because application code and input are available before execution, their size can be specified easily. Although output size is not available before task execution, in most cases it is a constant value with a known size or it can be estimated in terms of input value or input size.

### 4.2. Solver

If we suppose to offload either the whole task or nothing and every time we make decision for only one task, we can define the solver as a formal cost function. The cost function is calculated for the mobile device and every surrogate; either the mobile device or a surrogate with minimum cost value, would be the execution platform of the task.

In this paper, we suppose having context metrics to estimate the execution cost. We define the current processing power as Equation 1, where $P_u$ is the percentage of usage of processor, and $P_s$ is the processor speed.

$$P_c = (1 - P_u) \times P_s \qquad \textbf{(1)}$$

The cost function to determine the target execution location is defined by Equation 2.

$$Cost = \frac{w_1 * Time + w_2 * Energy}{w_3 * P_c + w_4 * Available\ Memory} \qquad \textbf{(2)}$$

Where $w_1$ to $w_4$ are the weighting factors which are non-negative values; the summation of them is one and represents the importance of the corresponding factors. Calculating the *Cost* for the mobile device, the *Time* factor is the execution time of the task on mobile device ($Time_{mobile}$) and the *Energy* factor referred to energy consumption at run-time that is defined as Equation 3.

$$Energy_{mobile} = Time_{mobile} \times Power_{Comp} \quad \textbf{(3)}$$

Energy consumption of mobile device in various states is different [20, 26]. Therefore, we have defined $Power_{comp}$ as power rate for computation on the mobile device. While $Power_{standby}$ is defined as power rate of mobile device on remote execution, and $Power_{send}$ and $Power_{receive}$ are power rate for sending and receiving data.

Calculating the *Cost* for surrogates, the *Time* factor is calculated by Equation 4, and the *Energy* factor is calculated by Equation 6.

$$Time_{surrogate} = Time_{send} + ExecutionTime_{surrogate} + Time_{receive} \quad \textbf{(4)}$$

$Time_{send}$ and $Time_{receive}$ are calculated in terms of *Transmission Data Size*, which includes the sizes of code, input data, and output data as given in Equation 5.

$$Time_{Send/receive} = \frac{Transmission\ Data\ Size}{Data\ Transmission\ Rate} \quad \textbf{(5)}$$

$$Energy_{surrogate} = (Time_{send} * Power_{Send}) + (ExecutionTime_{surrogate} * Power_{Standby}) + (Time_{receive} * Power_{receive}) \quad \textbf{(6)}$$

Figure 1 shows the pseudo code of our proposed solver.

```
Proposed_Solver()
{
  if ((Available_Memory_Mobile < Required_Memory_Application) or
      (Available_Energy_Mobile < Required_Energy_Application))
  {
    Mobile_In_Competition = FALSE;
  }
  foreach surrogate
  {
    Calculate time and energy to offload the task();
    if ((Available_Memory_Surrogate[i] < Required_Memory_Application) or
        (Available_Energy_Mobile< Required_Energy_Surrogate[i]))
    {
      Surrogate[i].In_Competition = FALSE;
    }
  }
  if (forall Surrogates: Surrogate[i].In_Competition == FALSE)
  {
    if (Mobile_In_Competition == TRUE)
      LocalExecution();
    else
      DoNothing();
  }
  else
  {
    foreach Surrogate/Mobile: if In_Competition == TRUE
    {
      CalculateCost;
    }
    Execute the task on the Surrogate/Mobile with minimal Cost();
  }
}
```

Figure 1. Solver algorithm

## 5. EVALUATION

To quantify the effectiveness of our proposed approach, we constructed a test bed consisting of one mobile device and one surrogate whose specifications are given in Table 1. The mobile device was connected to surrogates via 802.11b/g WLAN. The context information of mobile device, surrogate, and applications were presented in XML file format. Figure 2 shows the context information of mobile device, and Figure 3 shows the context information of the available surrogate in the chosen test bed.

Table 1. Configuration of devices used in our experimentations.

| **Type** | **Processor** | **Memory** | **Operating System** |
|---|---|---|---|
| Mobile | Qualcomm MSM7225™ 528 MHz | 256 MB | Windows Mobile 6.5 Professional |
| Surrogate | Intel Core 2Duo 2.5 GHz | 4 GB | Windows 7 Professional |

```
<MobileDevice>
  <NodeContext>
    <Name> Mobile </Name>
    <CPU> 524MHz </CPU>
    <InstructionPSecond> 270 </InstructionPSecond>
    <Load> 0.05 </Load>
    <AvailableMemory> 91MB </AvailableMemory>
    <AvailableBattery> 800J </AvailableBattery>
  </NodeContext>
</MobileDevice>
```

Figure 2. Context specification of the mobile device

```
<Surrogates>
  <NodeContext>
    <Name> Surrogate1 </Name>
    <CPU> 5000MHz </CPU>
    <InstructionPSecond> 938010 </InstructionPSecond>
    <Load> 0.1 </Load>
    <AvailableMemory> 2200MB </AvailableMemory>
    <Bandwidth> 1KB/S </Bandwidth>
  </NodeContext>
</Surrogates>
```

Figure 3. Context specification of the surrogates

We evaluated the effectiveness of our proposed approach with respect to responsiveness and resource consumption. We evaluated the responsiveness of the proposed approach through a scenario where the user intended to execute an application for finding the *n*th prime number, which is a CPU intensive application, and needed high computing power and low memory size on a mobile device where a surrogate was in range. Figure 4 shows the context information of the *n*th prime application.

We evaluated the proposed approach with respect to resource consumption through a scenario where the user intended to execute an application to determine a matrix determinate, which needed high computing power and size of input data was respectively high. Figure 5 shows the context information of matrix determinate application.

As we have stated earlier, in this paper we suppose context descriptor files for the mobile device, surrogates and tasks are prepared in advance. Furthermore, in both mentioned benched applications, output data has a constant size which is indicated by *BaseOutputSize* tag in Figure 4 and Figure 5. We measured the response time and resource consumption in three scenarios: local execution of application on mobile device, offloading the application and execution on surrogate, and using our proposed method to find the target execution location and run it.

```
<ApplicationContext>
  <Name> Nth Prime Number </Name>
  <RequiredMemory> 0.6MB </RequiredMemory>
  <CodeSize> 1KB </CodeSize
  <BaseInputSize> 0.05KB </BaseInputSize>
  <BaseOutputSize> 0.05KB </BaseOutputSiz>
  <Order>
    (N*ln(N)+(N*ln(ln(N))))*(pow(N*ln(N)+(N*ln(ln(N))),0.5))
  </Order>
</ApplicationContext>
```

Figure 4. Context specification of the *n*th prime number application

```
<ApplicationContext>
  <Name> Matrix Determinant </Name>
  <RequiredMemory> 9MB </RequiredMemory>
  <CodeSize> 2KB </CodeSize
  <BaseInputSize> 0.1KB </BaseInputSize>
  <BaseOutputSize> 0.05KB </BaseOutputSiz>
  <Order> N! </Order>
</ApplicationContext>
```

Figure 5. Context specification of the matrix determinant

## 5.1. Responsiveness

Responsiveness is defined as the time used by an application to respond to a user-triggered request [21]. In general, lower response time is more satisfactory and always we hope that response time is low enough for good subjective performance.

To estimate the execution time, we replace variable *N* with input value of the application in the function presented in the *Order* section of the application context description. The result is then divided by *InstructionPSecond* presented in context descriptor of mobile device and surrogate to estimate the execution time of the application on mobile device and surrogate.

Figure 6 shows the response time of the *n*th prime application among increases in input size. As it is shown, our proposed approach almost always yields the least response time and thus the best location to run the application.

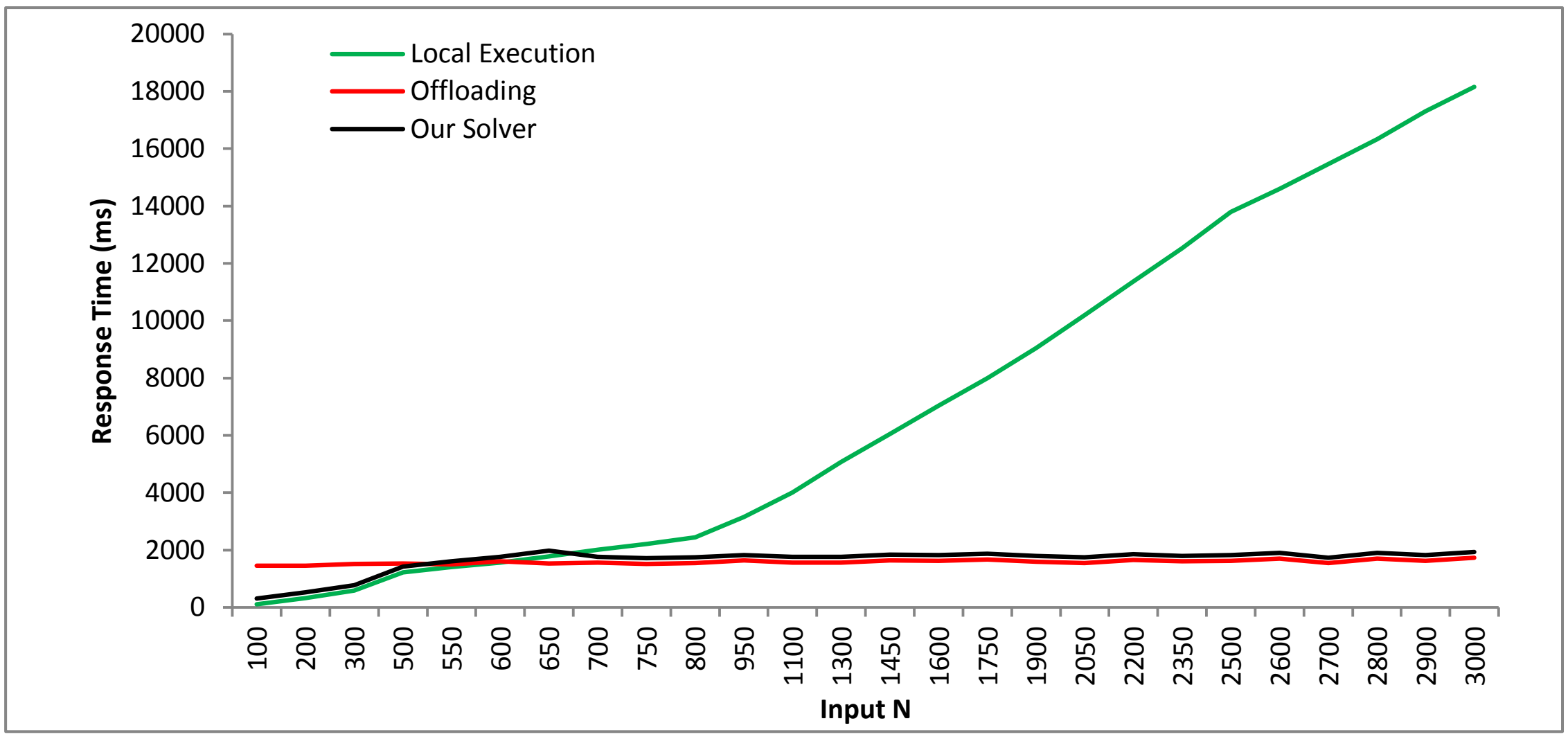

Figure 6. Comparison of execution time for finding the *n*th prime number

## 5.2. Energy

Battery lifetime is an important aspect of participating mobile devices in pervasive environments. Therefore, a good offloading mechanism should focus on consuming as low energy as possible. To evaluate the impact of cyber foraging on energy consumption, we experimentally measured the energy consumption of mobile devices through execution of a *matrix determinant* application.

In the *n*th prime application, a number with fixed size was the application's input data, but the *matrix determinant* application required to send the whole matrix to the surrogate. Therefore, the *Data Size* factor in Equation 5 was variable in terms of matrix's row count that affected the cost function and so the decision. In this scenario, due to simplicity, we assumed the $Power_{send}$, $Power_{receive}$, and $Power_{standby}$ factors as equal in Equation 6.

To emphasis on energy consumption in this scenario, we set the *Energy* weight ($w_2$) in Equation 2 to maximum value of 1. Figure 7 presents the energy consumption of execution of *matrix determinate* application; as it is shown, our proposed approach preserved the minimum energy consumption compared to local execution of the application in mobile device or always offloading the application and execution on the surrogate.

An issue that should be considered in every decision maker's mechanism is the execution overheads of the decision-making process itself, which must be as light as possible. As it is shown in Figure 6 and Figure 7, our proposed approach, preserved nearly the same response time and energy consumption compared to blind offloading approach, when it decides to offload the task; and nearly the same response time and energy consumption compared to local execution on mobile device, when it decides to execute the task on the mobile phone. Actually, the computational complexity of our proposed solver is O($n$) which $n$ is the number of available surrogates. Since the number of surrogates is always relatively small, the overhead of decision-making of our proposed solver does not affect the results and is almost negligible.

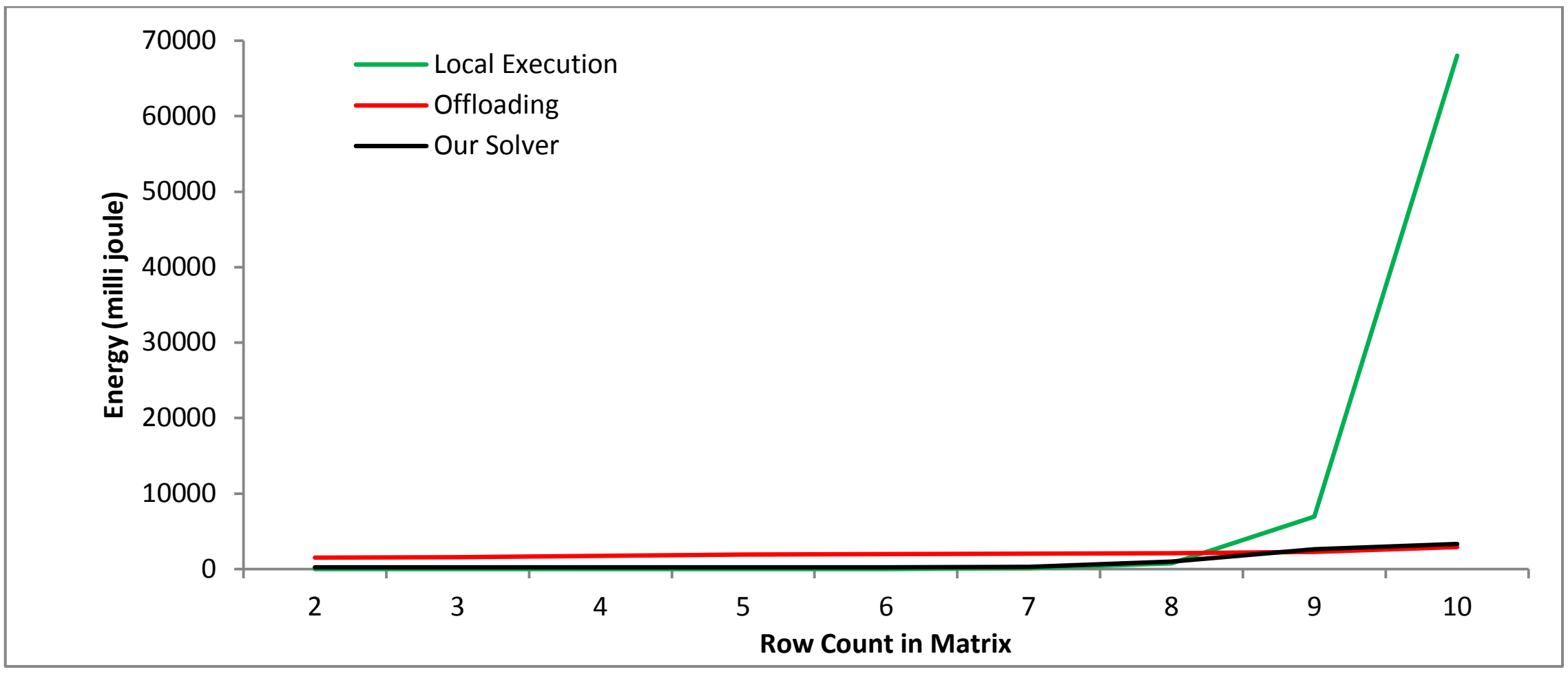

Figure 7. Comparison of energy consumption for *matrix determinant* application

## 6. Conclusion and Future Works

Mobile devices have always suffered from resource constraints, in comparison with static computers, to run complex and high computational applications. One of the major and most common solutions to improve computational resource poverty of mobile devices, especially in pervasive computing environments, is cyber foraging, which is offloading some tasks to more powerful nearby static computers. However, as discussed in this paper, cyber foraging is not effective in all circumstances and metrics such as mobile device and surrogate specifications, network quality, transmission data size, and application nature should be taken into account.

In this paper, we proposed a context-aware cyber foraging approach to ameliorate the resource poverty shortages of mobile devices and to raise the ability of participation of mobile devices in pervasive and mobile computing.

Experimental results showed the superiority of the proposed approach in response time and energy consumption, which are two most important metrics in mobile computing, in contrast to local execution of applications on mobile devices or blind offloading to surrogates.

As a future work, we are working on support of more than one surrogate and considering other metrics like surrogates' load, and surrogates' geographical distance that affects the wireless signal strength and network bandwidth. In addition, more experiments with various application types, and mobile/surrogates context, could increase the applicability of our proposed approach.